\documentclass[reqno]{amsart}

\usepackage[utf8]{inputenc}
\usepackage{xcolor}

\usepackage{amssymb}
\usepackage{mathtools}
\usepackage{amsthm}
\usepackage{bm}
\usepackage{hyperref}
\usepackage{autonum}
\usepackage{graphicx}
\usepackage{subcaption} 
\usepackage{pgfplots}
\usepgfplotslibrary{groupplots,dateplot}
\usetikzlibrary{patterns,shapes.arrows}
\usetikzlibrary{spy}
\usepackage{geometry}
\usepackage{siunitx}

\usepackage{algorithm}
\usepackage{algpseudocode}

\usepackage{hyperref}

\theoremstyle{definition}

\newtheorem{remark}{Remark}
\newtheorem{assumption}{Assumption}

\usepackage{siunitx}
\usepackage{pgfplots}
\pgfplotsset{compat=1.18}
\usepgfplotslibrary{groupplots}
\usepackage{standalone}
\pgfplotsset{
    lua backend=true,
}

\usepackage{robust-externalize}
\robExtConfigure{
add to preset={latex}{
use lualatex, 
},
add to preset={tikz}{
add to preamble={
              \usepackage{siunitx}
              \usepackage{pgfplots}

\newcommand{\signal}[1]{x_{#1}}                    
\newcommand{\signalvec}{\mathbf{x}}                
\newcommand{\estimateder}[1]{\hat{#1}}               
\newcommand{\signalderiv}[2]{\estimateder{x}_{#1}^{(#2)}} 
\newcommand{\baselinesig}[1]{\change{b_{#1}}}            
\newcommand{\baselinederiv}[2]{\change{\hat{b}_{#1}^{(#2)}}} 
\newcommand{\noisesig}[1]{\eta_{#1}}              
\newcommand{\noisederiv}[2]{\estimateder{\eta}_{#1}^{(#2)}} 
\newcommand{\powerderiv}[2]{\estimateder{h}_{#1}^{(#2)}} 

\newcommand{\timevar}{t}                           
\newcommand{\sampfreq}{f_{\mathrm{s}}}                        
\newcommand{\timewindow}{T}                        
\newcommand{\timedelay}[2]{\tau_{#1,#2}}          

\newcommand{\diffkernelsymbol}{g}                     
\newcommand{\diffkernel}[1]{\diffkernelsymbol^{(#1)}}               
\newcommand{\diffkernelshift}[2]{\diffkernelsymbol^{(#1)}_{#2}}  
\newcommand{\fourierkernel}{\fourierkernelsymbol}     
\newcommand{\fourierkernelsymbol}{\mathcal{G}}                     

\newcommand{\jacoparam}{\alpha}                    
\newcommand{\jacoparambeta}{\beta}                 
\newcommand{\polyorder}{N}                         
\newcommand{\jacopoly}[1]{P_{#1}^{(\jacoparam,\jacoparambeta)}} 
\newcommand{\jacobiweight}[1]{w(#1)}              
\newcommand{\derivorder}{n}                        
\newcommand{\filterlen}{L}                         

\newcommand{\pulseamp}[1]{a_{#1}}                  
\newcommand{\pulseamphat}[1]{\hat{a}_{#1}}        
\newcommand{\pulsetime}[2]{t_{#1}^{(#2)}}         
\newcommand{\arrivtime}[3]{t_{#1,#2}^{(#3)}}      
\newcommand{\numpulses}[1]{N_{#1}}                 
\newcommand{\windowset}[1]{\mathcal{W}_{#1}}      
\newcommand{\sourceset}{\mathcal{L}}               
\newcommand{\sourceHeart}{\text{Heart}}           
\newcommand{\sourceUterus}{\text{Uterus}}         

\newcommand{\chanindex}{i}                         
\newcommand{\numchannels}{m}                       

\newcommand{\dirac}{\delta}                        

\newcommand{\change}[1]{#1}
              \pgfplotsset{compat=1.18}
              \usepgfplotslibrary{groupplots}
              \pgfplotsset{table/search path={../}}},
              },
}
\cacheTikz

\title{\LARGE \bf
	Heart Artifact Removal in Electrohysterography Measurements Using Algebraic Differentiators
}

\author{Amine Othmane$^{1}$, Maria Camila Bustos Vivas$^{2}$, Johannes Steuer$^{1}$, Jana Hutter$^{3}$%
	\address{$^{1}$Amine Othmane and Johannes Steuer are with Systems Modeling and Simulation,
		Saarland University, Germany
		{\tt\small amine.othmane@uni-saarland.de}}%
	\address{$^{2}$Maria Camila Bustos Vivas is with the Radiology Department, Uniklinikum Erlangen, Germany
		{\tt\small maria.v.bustos@fau.de}}%
	\address{$^{3}$Jana Hutter is with the Institute for Information Processing, University of Hanover, Germany
		{\tt\small hutter@tnt.uni-hannover.de}}%
}

\thanks{*This work was not supported by any organization}%
\begin{document}
	
	\maketitle
	\thispagestyle{empty}
	
	\pagestyle{empty}

	\begin{abstract}
		Electrohysterography (EHG) enables non-invasive monitoring of uterine contractions but can be contaminated by electrocardiogram (ECG) artifacts. This work presents an ECG removal method using algebraic differentiators, a control-theoretic tool for model-free derivative estimation, that preserves signal shape outside the detected cardiac pulse locations. The differentiator parameters are designed to simultaneously suppress slow physiological artifacts and powerline interference while maximizing output signal-to-noise ratio. Cross-channel clustering distinguishes cardiac pulses from localized artifacts, enabling accurate pulse subtraction without auxiliary ECG references. Implemented as a causal FIR filter, the method is validated on multichannel EHG recordings from female and male subjects and compared to the template subtraction method.
	\end{abstract}
	
	\section{Introduction}
Electrohysterography (EHG) is a non-invasive method for recording uterine electrical activity associated with myometrial contractions \cite{Kuijsters2017UP, Garfield2007UC}, applied to pregnancy monitoring, preterm labor detection, and analysis of uterine peristalsis across the menstrual cycle \cite{Alberola-Rubio2026EH, Rabotti2015EHG, Wang2024EHGMRI}. Abnormal contractility has been linked to adenomyosis, endometriosis, and leiomyoma \cite{Kuijsters2017UP, Harmsen2023MRI_US_HI}, and EHG enables continuous non-invasive monitoring where invasive methods or imaging are impractical \cite{Harmsen2023MRI_US_HI, Garfield2007UC}. 

However, surface recordings of electrical activities  are contaminated by multiple physiological sources including electrocardiogram (ECG) and electromyographic activity from abdominal muscles \cite{abbaspour2014removing,Petersen2020,VANLEUTEREN2019176}.
ECG artifacts pose a challenge due to the potential large amplitude of cardiac signals \cite{abbaspour2014removing,Petersen2020}.
Conventional ECG removal methods have significant limitations. QRS interpolation methods risk signal loss within the excised window~\cite{STAM2023147,VANLEUTEREN2019176}.
Adaptive filtering requires auxiliary ECG reference channels \cite{abbaspour2014removing}. Template subtraction assumes stationary ECG morphology and can introduce residual artifacts when the ECG waveform varies \cite{Petersen2020}, while wavelet-based methods demand substantial computational resources or remove some low spectral components of the measured signals~\cite{Miljković2017RemovingECG}.

This work addresses ECG artifact removal using algebraic differentiators, a tool from control theory for model-free derivative estimation \cite{Mboup2009, Othmane2022}, well-suited for detecting sparse impulsive cardiac events in signals corrupted by baseline drift and periodic interference, with noise rejection and low computational complexity via causal FIR filtering \cite{kiltz2014, Kiltz2013}.

The main contributions are: (i) a distributional signal model that represents impulsive physiological events, (ii) a systematic cardiac artifact removal method based on algebraic differentiators with constraint-based parameter selection for simultaneous powerline interference suppression and baseline attenuation, including theoretical analysis of noise rejection and frequency selectivity, (iii) a cross-channel clustering algorithm exploiting spatial coherence to distinguish cardiac pulses from localized artifacts without auxiliary ECG references, and (iv) \change{proof-of-concept validation on EHG recordings from two healthy volunteers, comparing the proposed method against template subtraction and demonstrating effective artifact removal that preserves signal integrity outside the detected pulse locations, with low computational complexity and no training requirements.}

The paper is structured as follows: Section~\ref{sec:data} describes the data acquisition. Section~\ref{sec:model} presents the signal model. The removal algorithm is detailed in Section~\ref{sec:artifact_removal} and  Section~\ref{sec:results} presents experimental results. Section~\ref{sec:conclusion} concludes the works and presents discusses future works.
	\section{Data Acquisition and Problem Statement}\label{sec:data}
\subsection{Data Acquisition}
\begin{figure}
     \centering
     \begin{subfigure}[b]{0.5\textwidth}
         \centering
         \includegraphics[width=0.8\textwidth]{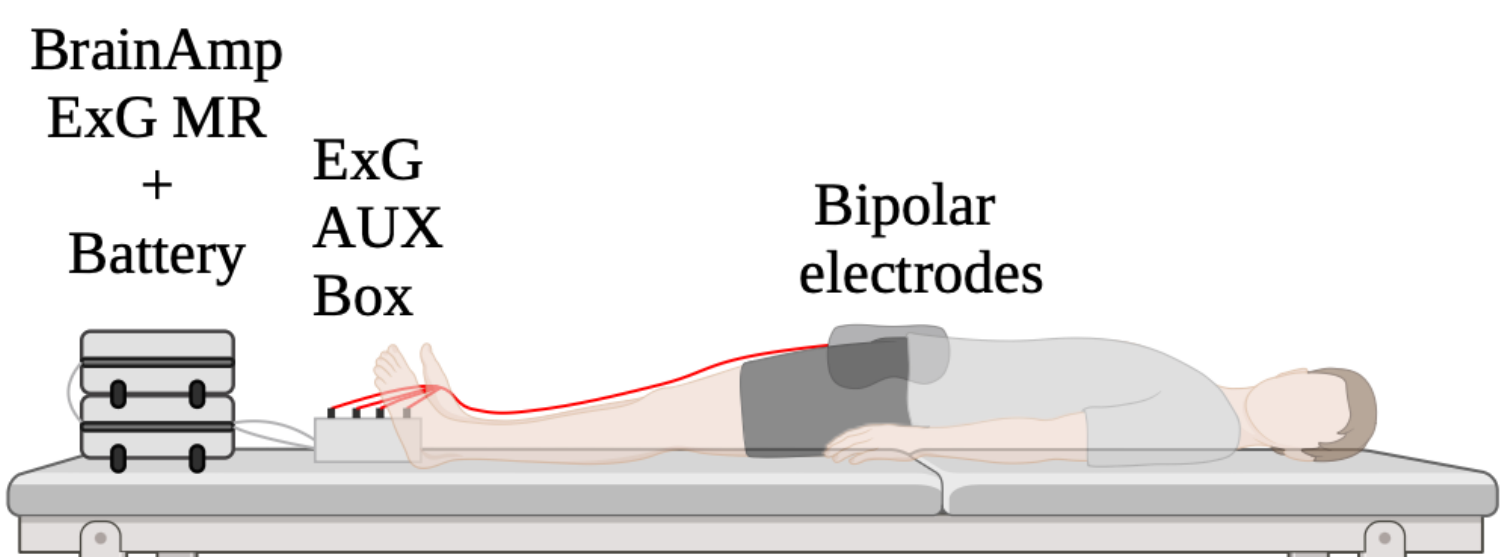}
         \caption{Experimental setup for EHG signal acquisition.}
         \label{fig:SetUp}
         \hspace{1mm}
     \end{subfigure}
     \begin{subfigure}[b]{0.45\textwidth}
         \centering
         \includegraphics[width=0.8\textwidth]{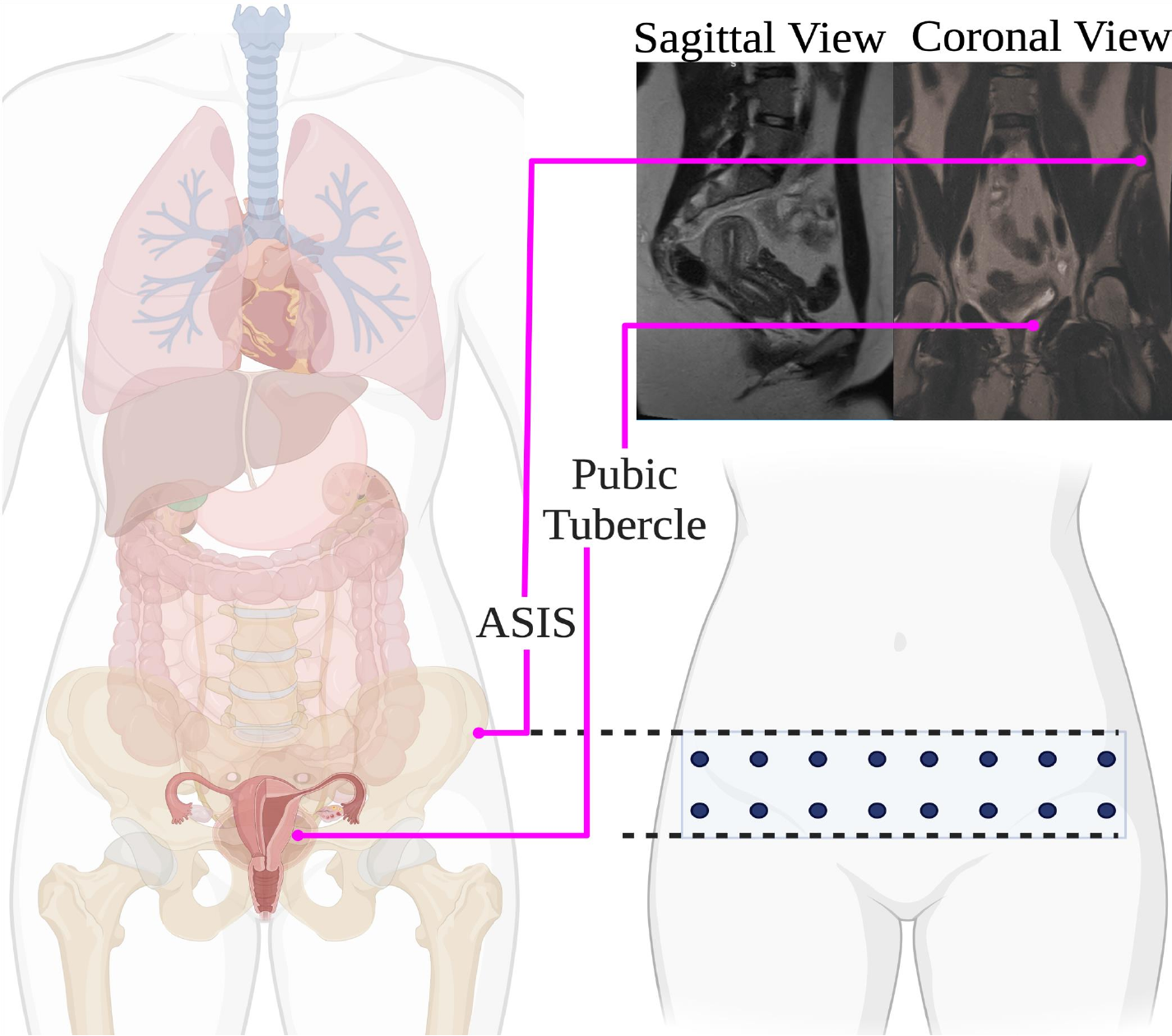}
         \caption{Electrode placement strategy.}
         \label{fig:ElectrodePlacement}
     \end{subfigure}
        \caption{(a) EHG recording setup. (b) Electrode placement using the anterosuperior iliac spine (ASIS) and pubic bone as anatomical landmarks, shown on sagittal and coronal MRI scans.}
        \label{fig:DataAcquisition}
\end{figure}

The study population comprised two healthy volunteers, one non-pregnant woman (aged 21–30 years) and one healthy male subject (aged 24–30 years) to identify uterine-specific signal components absent in male recordings. Participants lay supine with minimal movement (Figure~\ref{fig:SetUp}). \change{Skin preparation with abrasive gel reduced electrode impedance below $\SI{10}{\ohm}$.} Electrodes were placed between the anterosuperior iliac spine (ASIS) and pubic bone, with a ground electrode on one hip (Figure~\ref{fig:ElectrodePlacement}). EHG signals were recorded using a multichannel BrainAmp ExG MR amplifier (Brain Products, Gilching, Germany) with eight bipolar electrodes at a sampling frequency of $\sampfreq=\SI{5}{\kHz}$ for at least seven minutes per subject.

\subsection{Problem Statement}
The goal is to extract uterine-specific signal components from multichannel EHG recordings contaminated by ECG artifacts. Formally, given an observed signal, we seek to estimate the artifact-free signal by detecting and removing cardiac pulse contributions while preserving uterine activity, baseline physiological components, and minimizing signal distortion. The method should operate without auxiliary ECG reference channels and be suitable for real-time processing.

	\section{Mathematical Model}\label{sec:model}

Abdominal surface electromyography captures the superimposed electrical activity of multiple physiological sources. We consider an $\numchannels$-channel EHG signal $\signalvec(\timevar) = [\signal{1}(\timevar), \ldots, \signal{\numchannels}(\timevar)]^T \in \mathbb{R}^{\numchannels}$ observed over $\timevar \in [0, T_{\text{obs}}]$ and sampled at frequency $\sampfreq$.

\subsection{Signal Model}

Each physiological source can generate discrete events (e.g., heartbeats, uterine contractions) that propagate through the abdominal tissue via volume conduction and are detected as EHG signals. The primary objective is to decompose the observed multi-channel signal into a cardiac component consisting of impulse trains from heartbeats, a uterine component, and a residual baseline containing respiratory artifacts and other unmodeled sources.

The observed EHG signal in each channel $\chanindex\in\{1,\ldots,\numchannels\}$ is modeled as a superposition of physiological impulse trains from multiple organs, a baseline signal satisfying the following regularity condition, and deterministic power supply interference.

\change{\begin{assumption}[Locally Smooth Baseline]
\label{ass:smooth_baseline}
For each channel $\chanindex \in \{1,\ldots,\numchannels\}$, the baseline signal $\baselinesig{\chanindex} \in C^{\derivorder}([0, T_{\text{obs}}])$ is at least $\derivorder$ times continuously differentiable, where $\derivorder \in \mathbb{N}$ is the derivative order. Moreover, there exists $\epsilon > 0$ such that
$\sup_{t\in[0, T_{\text{obs}}]}|\baselinesig{\chanindex}^{(\derivorder)}(t)| < \epsilon,$
where $\epsilon$ is small relative to the peak magnitudes of the $\derivorder$-th derivatives of the impulsive components. This enables separation of impulsive events from the slowly-varying baseline via higher-order differentiation: $\baselinesig{\chanindex}$ can be locally approximated by polynomials of degree less than $\derivorder$ with error $\mathcal{O}(T_w^{\derivorder+1})$ within each window of length $T_w$.
\end{assumption}}

In the distributional sense, we have
\begin{equation}
\signal{\chanindex}(\timevar) = \baselinesig{\chanindex}(\timevar) + \sum_{\ell \in \sourceset} \sum_{j=1}^{\numpulses{\ell}} \pulseamp{\ell}^{(j)} \, \dirac(\timevar - \pulsetime{\ell}{j} - \timedelay{\chanindex}{\ell}) + h_{\chanindex}(\timevar) + \noisesig{\chanindex}(\timevar),
\label{eq:signal_model_dirac}
\end{equation}
where ${\baselinesig{\chanindex}: [0, T_{\text{obs}}] \to \mathbb{R}}$ represents slow-varying components satisfying Assumption~\ref{ass:smooth_baseline}, $\change{\sourceset = \{\sourceHeart, \sourceUterus\}}$ is the set of physiological sources, $\numpulses{\ell} \in \mathbb{N}_0$ is the number of events from source $\ell$ during the observation period, $\pulseamp{\ell}^{(j)} \in \mathbb{R}$ is the amplitude of the $j$-th impulse from source $\ell$, $\dirac$ denotes the Dirac delta distribution, $\pulsetime{\ell}{j} \in (0, T_{\text{obs}})$ is the occurrence time of the $j$-th event at source $\ell$, $\timedelay{\chanindex}{\ell} \geq 0$ is the propagation delay from source $\ell$ to channel $\chanindex$ (accounting for finite signal propagation velocity through tissue and spatial sensor distribution), $h_{\chanindex}: [0, T_{\text{obs}}] \to \mathbb{R}$ represents the power supply interference, and $\noisesig{\chanindex}$ represents zero-mean additive measurement noise.

The power supply interference $h_{\chanindex}$ is modeled as
$h_{\chanindex}(\timevar) = \sum_{q=1}^{Q} A_{\chanindex,q} \sin(2\pi q f_0 \timevar + \phi_{\chanindex,q})$,
where $f_0 = \SI{50}{\Hz}$ is the fundamental power line frequency, $Q \in \mathbb{N}$ is the number of harmonics, $A_{\chanindex,q} \geq 0$ are the harmonic amplitudes, and $\phi_{\chanindex,q} \in [0, 2\pi)$ are the channel-dependent phase offsets.

\subsection{Model Discussion}

The model \eqref{eq:signal_model_dirac} relies on two key idealizations: the Dirac delta representation of impulsive events and the smoothness of the baseline signal. Both are motivated by temporal scale separation inherent in EHG signals. Physiologically, cardiac R-waves (10--20~ms) and uterine electromyographic bursts are short relative to baseline variations from respiratory modulation and postural drift, so impulsive events appear instantaneous relative to baseline dynamics, justifying the Dirac representation.

\section{Artifact Removal via Pulse Detection and Reconstruction}\label{sec:artifact_removal}
This section presents a systematic approach to cardiac artifact removal that exploits the impulsive nature of cardiac pulses. Algebraic differentiation with carefully chosen parameters transforms the signal model \eqref{eq:signal_model_dirac} into a form where impulsive components dominate over baseline drift, power line interference, and measurement noise. Threshold-based detection identifies pulse start times in each channel, and cross-channel clustering distinguishes cardiac pulses from localized artifacts.
\subsection{Pulse Indicator Design}

To enhance impulse detection, we compute the $\derivorder$-th order derivative of the signal $\signal{\chanindex}$ in channel $\chanindex \in \{1,\ldots,\numchannels\}$ using algebraic differentiation \cite{Mboup2009,Othmane2022}. The $\derivorder$-th order time derivative of $\signal{\chanindex}$ is approximated by $\signalderiv{\chanindex}{\derivorder}(\timevar)$ defined as
\begin{equation}
\signalderiv{\chanindex}{\derivorder}(\timevar) = \int_{\timevar-\timewindow}^{\timevar} \diffkernel{\derivorder}(\timevar-\tau)\signal{\chanindex}(\tau)\,d\tau,
\label{eq:differentiation_operation}
\end{equation}
where $\diffkernel{\derivorder}$ is the $\derivorder$-th order derivative of a causal kernel  $\diffkernelsymbol$ with compact support $[0,\timewindow]$ (with $\timewindow$ in seconds). The kernel is constructed from weighted Jacobi polynomials $\jacopoly{j}(\nu)$ for $j = 0, 1, \ldots, \polyorder$, which form an orthogonal basis with respect to the weight function 
\begin{equation}
    \jacobiweight{\nu} = \begin{cases}
        (1-\nu)^{\jacoparam}(1+\nu)^{\jacoparambeta}, & \nu \in [-1,1], \\
        0, & \text{otherwise},
    \end{cases}
\end{equation}
for $\jacoparam, \jacoparambeta > -1$. The affine transformation $\nu(\tau) = 1 - 2\tau/\timewindow$ maps $[0,\timewindow]$ to $[-1,1]$. Truncating the polynomial expansion at order $\polyorder \in \mathbb{N}_0$ yields
\begin{equation}
\diffkernelsymbol(\tau)=\frac{2}{\timewindow}\,\jacobiweight{\nu(\tau)}\sum_{j=0}^{\polyorder}\frac{\jacopoly{j}(\vartheta)}{\|\jacopoly{j}\|_{L^2_{w}}^2}\jacopoly{j}(\nu(\tau)),
\end{equation}
where ${\vartheta \in [-1,1]}$ is a delay parameter and $$\|\jacopoly{j}\|_{L^2_{w}}^2 \coloneq\int_{-1}^{1} |\jacopoly{j}(\nu)|^2 \jacobiweight{\nu} \, d\nu$$ is the weighted $L^2$-norm. Higher-order derivative kernels $\diffkernel{\derivorder}$ are obtained by successive differentiation.

Applying the differentiation operation \eqref{eq:differentiation_operation} to the signal model \eqref{eq:signal_model_dirac}, the derivative signal becomes
\begin{equation}
\signalderiv{\chanindex}{\derivorder}(\timevar) = \signalderiv{\chanindex,\text{pulse}}{\derivorder}(\timevar) + \baselinederiv{\chanindex}{\derivorder}(\timevar) + \powerderiv{\chanindex}{\derivorder}(\timevar) + \noisederiv{\chanindex}{\derivorder}(\timevar),
\label{eq:derivative_signal_model}
\end{equation}
where $\signalderiv{\chanindex,\text{pulse}}{\derivorder}(\timevar) = \sum_{\ell,j} \pulseamp{\ell}^{(j)} \diffkernel{\derivorder}(\timevar - \arrivtime{\chanindex}{\ell}{j})$,  $\arrivtime{\chanindex}{\ell}{j} = \pulsetime{\ell}{j} + \timedelay{\chanindex}{\ell}$ denotes the arrival time of the $j$-th pulse from source $\ell$ at channel $\chanindex$, and $\baselinederiv{\chanindex}{\derivorder}$, $\powerderiv{\chanindex}{\derivorder}$, $\noisederiv{\chanindex}{\derivorder}$ are the $\derivorder$-th derivatives of the baseline, power supply interference, and noise, respectively. By the sifting property of the Dirac delta, each impulse $\dirac(\timevar - \arrivtime{\chanindex}{\ell}{j})$ is transformed under convolution into the shifted kernel $\diffkernel{\derivorder}(\timevar - \arrivtime{\chanindex}{\ell}{j})$, providing a characteristic signature for detection.

\subsection{Parametrizations for Artifact Suppression}

For the special choice $\polyorder=0$ and symmetric parameters ${\jacoparam = \jacoparambeta}$, the Fourier transform of $\diffkernelsymbol$ admits infinitely many zeros corresponding to the zeros of the Bessel function $J_{\jacoparam+1/2}$ of the first kind.

\subsubsection*{Power Interference Suppression}

Let $j_{\jacoparam+1/2,k} > 0$ denote the $k$-th positive zero of $J_{\jacoparam+1/2}$. For line frequency $f_0=\SI{50}{\Hz}$, set $\omega_0 = 2\pi f_0$ and choose
\begin{equation}
\timewindow = {2 j_{\jacoparam+1/2, k}}/{\omega_0}.
\label{eq:window_choice}
\end{equation}
This yields $\fourierkernel(\omega_0) = 0$, annihilating the fundamental harmonic. Since $|\fourierkernel(\omega)| = \mathcal{O}(|\omega|^{-(\jacoparam+1)})$ \cite{Kiltz2013} and differentiation introduces $|\omega|^{\derivorder}$, the power interference at harmonics $\omega_q = q\omega_0$ satisfies
$
|\mathcal{F}\{\powerderiv{\chanindex}{\derivorder}\}(\omega_q)| = \mathcal{O}(q^{\derivorder-\jacoparam-1})$, for $q \to \infty$.
Choosing $\jacoparam > \derivorder - 1$ ensures asymptotic decay for all harmonics $q \geq 2$.

\subsubsection*{Noise Amplification Analysis}

Assume $\noisesig{\chanindex}$ is a zero-mean wide-sense stationary process with power spectral density $S_{\noisesig{\chanindex}}$. The convolution \eqref{eq:differentiation_operation} transforms the power spectral density as
$
S_{\noisederiv{\chanindex}{\derivorder}}(\omega) = |\omega|^{2\derivorder} |\fourierkernel(\omega)|^2 S_{\noisesig{\chanindex}}(\omega).
$
For white noise with $S_{\noisesig{\chanindex}}(\omega) = S_0$, the variance satisfies
\begin{equation}
\sigma_{\derivorder}^2 := \mathrm{Var}[\noisederiv{\chanindex}{\derivorder}(\timevar)] = \frac{S_0}{2\pi} \int_{-\infty}^{\infty} |\omega|^{2\derivorder} |\fourierkernel(\omega)|^2 \, d\omega.
\label{eq:noise_variance}
\end{equation}
Since the kernel $\diffkernelsymbol$ has compact support $[0,\timewindow]$, this integral is always finite. However, since $|\fourierkernel(\omega)|^2 = \mathcal{O}(|\omega|^{-2(\jacoparam+1)})$, the noise amplification scales as $\mathcal{O}(|\omega|^{2\derivorder - 2(\jacoparam+1)})$ at high frequencies. Larger values of $\jacoparam$ relative to $\derivorder$ reduce noise amplification by accelerating the asymptotic decay of the kernel's frequency response \cite{Kiltz2013,mboup2018,Othmane2022}.

\subsubsection*{Detectability Condition}

Define $A_{\text{peak}} := \min_{\ell,j} |\pulseamp{\ell}^{(j)}| \cdot \|\diffkernel{\derivorder}\|_\infty$ as the minimum peak impulsive magnitude.
By Assumption~\ref{ass:smooth_baseline} with $\|\baselinederiv{\chanindex}{\derivorder}\|_\infty < \epsilon$, and the preceding analysis, $\signalderiv{\chanindex}{\derivorder}$ serves as an effective pulse indicator when
\begin{equation}
\max\{\epsilon, \, \|\powerderiv{\chanindex}{\derivorder}\|_\infty, \, \|\noisederiv{\chanindex}{\derivorder}\|_\infty\} \ll A_{\text{peak}}.
\label{eq:detectability_condition}
\end{equation}
The parameters $(\derivorder, \jacoparam, k)$ satisfying $\jacoparam > \derivorder - 1$ and~\eqref{eq:window_choice} are selected from the resulting feasible set on representative data; the values used in the experiments are reported in Section~\ref{sec:results}.

\subsection{Threshold-Based Pulse Start Time Detection}

The derivative signal \eqref{eq:derivative_signal_model} decomposes as
\begin{equation}
\signalderiv{\chanindex}{\derivorder}(\timevar) = \signalderiv{\chanindex,\text{pulse}}{\derivorder}(\timevar) + \signalderiv{\chanindex,\text{artifact}}{\derivorder}(\timevar),
\end{equation}
where $\signalderiv{\chanindex,\text{artifact}}{\derivorder} := \baselinederiv{\chanindex}{\derivorder} + \powerderiv{\chanindex}{\derivorder} + \noisederiv{\chanindex}{\derivorder}$ combines all artifact components. Under the detectability condition \eqref{eq:detectability_condition}, peaks in $|\signalderiv{\chanindex}{\derivorder}(\timevar)|$ are dominated by impulsive components.

For each channel $\chanindex \in \{1,\ldots,\numchannels\}$, let $\{|\signalderiv{\chanindex}{\derivorder}[\change{s}]| : \change{s} = 1, \ldots, K\}$, where $x[\change{s}] := x(\change{s}/\sampfreq)$ denotes the sampled signal, be the magnitude sequence over the observation window. Define $\tau_p^{(\chanindex)}$ as the $p$-th percentile of this empirical distribution. Candidate impulse times are identified at sample indices $\change{s}$ satisfying $|\signalderiv{\chanindex}{\derivorder}[\change{s}]| > \tau_p^{(\chanindex)}$. Let $\mathcal{T}_{\chanindex}^{(\derivorder)} = \{\change{s}_1^{(\chanindex)}, \ldots, \change{s}_{M_{\chanindex}}^{(\chanindex)}\}$ with $\change{s}_1^{(\chanindex)} < \cdots < \change{s}_{M_{\chanindex}}^{(\chanindex)}$ denote these ordered threshold crossings.

To avoid multiple detections from a single impulse, crossings within $L_{\min} = \lfloor \timewindow \sampfreq \rfloor$ samples are consolidated. The filtered detection set $\mathcal{D}_{\chanindex}^{(\derivorder)} \subseteq \mathcal{T}_{\chanindex}^{(\derivorder)}$ is constructed iteratively: initialize $\mathcal{D}_{\chanindex}^{(\derivorder)} = \{\change{s}_1^{(\chanindex)}\}$, then for $j = 2, \ldots, M_{\chanindex}$, add $\change{s}_j^{(\chanindex)}$ to $\mathcal{D}_{\chanindex}^{(\derivorder)}$ if and only if $\change{s}_j^{(\chanindex)} - \max \mathcal{D}_{\chanindex}^{(\derivorder)} \geq L_{\min}$. This consolidation exploits the compact support $[0,\timewindow]$ of $\diffkernel{\derivorder}$: a single impulse cannot generate threshold crossings separated by more than $\lfloor \timewindow \sampfreq \rfloor$ samples.

For each detection ${\change{s}_j^{(\chanindex)} \in \mathcal{D}_{\chanindex}^{(\derivorder)}}$, we estimate the pulse start time at channel $\chanindex$ from the first zero-crossing $z_j^{(\chanindex)}$ of $\signalderiv{\chanindex}{\derivorder}$ following the peak magnitude at $\change{s}_j^{(\chanindex)}$, which is more robust to noise than preceding zero-crossings. \change{The post-peak zero-crossing is preferred because the leading edge of the kernel signature may be contaminated by the trailing edge of a preceding pulse or by transient baseline components, whereas the post-peak zero-crossing occurs within the well-isolated main lobe of $\diffkernel{\derivorder}$ and its location is insensitive to amplitude scaling.}

\subsection{Cross-Channel Cardiac Pulse Clustering}

Cardiac events propagate across all channels with small inter-channel delays, while localized artifacts (bowel movements, motion) affect fewer channels. This spatial coherence enables unsupervised cardiac pulse identification by clustering detections from all channels $\chanindex = 1, \ldots, \numchannels$.

Collect all detected pulses into $\mathcal{P} = \{(t_j^{(\chanindex)}, \chanindex, j) : \chanindex \in \{1,\ldots,\numchannels\}, \, j = 1, \ldots, |\mathcal{D}_{\chanindex}^{(\derivorder)}|\}$ and sort by time. Group pulses into candidate beat groups $\change{\mathcal{B}_{m_g}}$ ($\change{m_g = 1, 2, \ldots}$) using temporal window $\Delta t_{\text{beat}} \in [20, 50]$ ms: initialize $\mathcal{B}_1$ with the earliest pulse, then for each subsequent pulse $(t, \chanindex, j) \in \mathcal{P}$, add it to the current group $\change{\mathcal{B}_{m_g}}$ if $t - \max\{t' : (t', \cdot, \cdot) \in \change{\mathcal{B}_{m_g}}\} \leq \Delta t_{\text{beat}}$, otherwise start a new group $\change{\mathcal{B}_{m_g+1}}$. Classify $\change{\mathcal{B}_{m_g}}$ as cardiac if $N_{\text{ch}}(\change{\mathcal{B}_{m_g}}) = |\{\chanindex : (\cdot, \chanindex, \cdot) \in \change{\mathcal{B}_{m_g}}\}| \geq \lceil \rho \cdot \numchannels \rceil$ with $\rho \in [0.5, 0.7]$; otherwise mark as artifact. Within each cardiac beat group $\change{\mathcal{B}_{m_g}}$, compute median time $\change{\bar{t}_{m_g}} = \text{median}\{t : (t, \cdot, \cdot) \in \change{\mathcal{B}_{m_g}}\}$ and retain per channel only the detection $(t_j^{(\chanindex)}, \chanindex, j)$ minimizing $|t_j^{(\chanindex)} - \change{\bar{t}_{m_g}}|$. \change{The window $\Delta t_{\text{beat}}$ should exceed the maximum expected inter-channel propagation delay of the cardiac wavefront; values in $[20, 50]$~ms cover the physiological range of volume-conducted cardiac delays for standard abdominal electrode spacings, consistent with the inter-channel delay spread observed in Figure~\ref{fig:boxplot_delays}. The threshold $\rho \in [0.5, 0.7]$ requires simultaneous detection in at least half the channels, exploiting the global nature of cardiac propagation while tolerating occasional missed detections in low-SNR channels.}

\begin{remark}[Algorithm Limitations]
The greedy grouping may merge nearby beats when inter-beat intervals approach $2\Delta t_{\text{beat}}$, and ectopic beats or motion artifacts confined to a subset of channels may be misclassified as localized artifacts depending on $\rho$. The hard channel participation threshold furthermore lacks adaptivity to varying SNR conditions. 

\change{As a proof-of-concept study, these corner cases have not been systematically evaluated; their impact on detection performance should be assessed in future validation on larger and more diverse cohorts.} \change{Additionally, the percentile threshold is computed globally over the full observation window. For non-stationary recordings with time-varying uterine or motion activity, a sliding-window approach would be more appropriate; the global threshold is appropriate for the quasi-stationary resting conditions of the current recordings.}
\end{remark}

\subsection{Amplitude Estimation and Signal Reconstruction}

Following cardiac pulse clustering, let $\mathcal{C}_{\chanindex} \subseteq \mathcal{D}_{\chanindex}^{(\derivorder)}$ denote the subset of detections classified as cardiac pulses in channel $\chanindex$, with $|\mathcal{C}_{\chanindex}| = N_{\chanindex}$ cardiac pulses. For the $m$-th cardiac detection ($m = 1, \ldots, N_{\chanindex}$) with zero-crossing $z_m^{(\chanindex)}$, define $t_m := z_m^{(\chanindex)} \sampfreq^{-1}$ and the time-shifted kernel $\diffkernelshift{\derivorder}{m}(\timevar) = \diffkernel{\derivorder}(\timevar - t_m + \timevar_0)$, where $\timevar_0$ is the post-peak zero-crossing of $\diffkernel{\derivorder}(\timevar)$. The amplitude $\pulseamphat{m}^{(\chanindex)}$ is estimated via least squares over $\windowset{m} = [t_m, \, t_m + \timewindow]$, yielding
\begin{equation}
\pulseamphat{m}^{(\chanindex)} = \frac{\sum_{\change{s} \in \mathcal{I}_m} \signalderiv{\chanindex}{\derivorder}[\change{s}]  \diffkernelshift{\derivorder}{m}[\change{s}]}{\sum_{\change{s} \in \mathcal{I}_m} \left( \diffkernelshift{\derivorder}{m}[\change{s}] \right)^2},
\label{eq:amplitude_ls}
\end{equation}
where $\mathcal{I}_m$ denotes sample indices in $\windowset{m}$. The output signal $\signal{\chanindex,\text{clean}}$ is defined as
$$
\signal{\chanindex,\text{clean}}(\timevar) = \int_{\timevar-\timewindow}^{\timevar} \diffkernelsymbol(\timevar-\tau)\signal{\chanindex}(\tau)\,d\tau - \signalderiv{\chanindex,\text{pulse}}{0}(\timevar),
$$
 and the cardiac pulse component is reconstructed as
$
    \signalderiv{\chanindex,\text{pulse}}{0}=\sum_{m=1}^{N_{\chanindex}}\pulseamphat{m}^{(\chanindex)} \diffkernelshift{0}{m}(\timevar),
$ where $\diffkernelshift{0}{m}(\timevar) = \diffkernelsymbol(\timevar - t_m + \timevar_0)$.
\change{The signal $\signal{\chanindex,\text{clean}}$ is thus not the original signal with cardiac pulses excised, but the output of the zeroth-order differentiator applied to $\signal{\chanindex}$, with the reconstructed cardiac component subtracted. The zeroth-order kernel $\diffkernelsymbol$ used in the next section acts as a causal low-pass filter with cutoff at $\omega_c \approx 82.4$~rad/s, attenuating high-frequency noise and power-line interference while preserving the EHG frequency band of interest.}

	\section{Experimental Results}\label{sec:results}
Approximating \eqref{eq:differentiation_operation} with the mid-point rule yields a causal FIR filter of length $\filterlen = \lceil \timewindow \sampfreq \rceil$ samples \cite{Othmane2022}. The results use parameters $\derivorder=3$, ${\jacoparam = \jacoparambeta = 12}$, $\polyorder=0$, and $k=6$ (6th Bessel zero), yielding from \eqref{eq:window_choice} the window length $\timewindow \approx \SI{0.226}{\second}$ and $\filterlen = 1129$ at $\sampfreq = \SI{5}{\kHz}$. \change{The midpoint-rule discretization preserves the 50 Hz null: the magnitude of the FIR filter frequency response at $\SI{50}{\Hz}$ is approximately $10^{-17}$, confirming that discretization error is negligible.} \change{The detection and clustering parameters are $p = 97$, $\Delta t_{\text{beat}} = \SI{30}{\ms}$, and $\rho = 0.5$.} To characterize removal quality, the output SNR is defined as
\begin{equation}
\text{SNR}_{\text{out}} \coloneq \frac{\|\signalderiv{\chanindex,\text{pulse}}{\derivorder}\|_{L^2}^2}{\|\signalderiv{\chanindex}{\derivorder} - \signalderiv{\chanindex,\text{pulse}}{\derivorder}\|_{L^2}^2},
\label{eq:SNR}
\end{equation}
where $\signalderiv{\chanindex,\text{pulse}}{\derivorder}$ is the reconstructed cardiac pulse component; the denominator quantifies residual energy not accounted for by detected pulses, requiring no ECG ground truth.

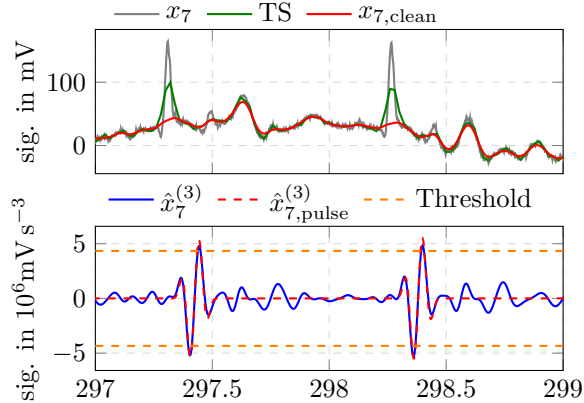
\begin{figure}
    \centering
    \begin{tikzpicture}[]
\pgfplotsset{
    ylabel with exponent/.style n args={3}{
        ylabel={#1 #3},
        scaled y ticks=true,
        ytick scale label code/.code={%
            \pgfmathparse{int(##1)}%
            \xdef#2{\pgfmathresult}%
        },
        after end axis/.code={%
            \ifx#2\undefined\xdef#2{0}\fi
            \pgfmathparse{#2}%
            \ifnum\pgfmathresult=0\relax
                \node[anchor=south,rotate=90] at (rel axis cs:-0.1,0.5) {#1 #3};
            \else
                \node[anchor=south,rotate=90] at (rel axis cs:-0.1,0.5) {#1 $10^{#2}$#3};
            \fi
        },
        ylabel style={opacity=0},
        y tick scale label style={opacity=0},
    }
}

    \def\datafileA{figs/experimental_results/csv_files_and_hash_info_male/csv_files_and_hash_info_male/channel_7_data_row0_col0.dat}
    \def\datafileB{figs/experimental_results/csv_files_and_hash_info_male/csv_files_and_hash_info_male/channel_7_data_row1_col0.dat}
    \def\datafileC{figs/experimental_results/csv_files_and_hash_info_male/csv_files_and_hash_info_male/channel_7_data_row2_col0.dat}
    \def\datafileD{figs/experimental_results/csv_files_and_hash_info_male/csv_files_and_hash_info_male/channel_7_data_row1_col0_thresholds.dat}
    \def\datafileTS{figs/experimental_results/csv_files_and_hash_info_male/csv_files_and_hash_info_male/202509010002_epi_channel_7_ecg_removed_TS.dat}


\pgfplotstableread{\datafileB}\datatable
\pgfplotstablegetrowsof{\datatable}
\pgfmathtruncatemacro{\lastrow}{\pgfplotsretval-1}
\pgfplotstablegetelem{0}{time}\of{\datatable}
\pgfmathsetmacro{\xminval}{\pgfplotsretval}
\pgfplotstablegetelem{\lastrow}{time}\of{\datatable}
\pgfmathsetmacro{\xmaxval}{\pgfplotsretval}

    \begin{groupplot}[
        group style={
            group size=1 by 2,
            vertical sep=0.7cm,
            xlabels at=edge bottom,
            xticklabels at=edge bottom,
        },
        width=7.8cm,
        height=3.5cm,
        legend pos=north east,
        legend columns=3,
        grid=major,
        grid style={dashed,gray!30},
        legend style={draw=none, at={(0.75,1.05)}, fill=none, anchor=north east, inner sep=1pt, outer sep=0pt, /tikz/every even column/.append style={column sep=3pt}},
        cycle list name=color list,
        scaled y ticks=false,
        every y tick scale label/.style={at={(yticklabel* cs:1.35)},anchor=south},
        xmin=297, xmax=299,
    ]
    
    \nextgroupplot[legend style={draw=none, at={(0.75,1.24)}, fill=none, anchor=north east, inner sep=1pt, outer sep=0pt, /tikz/every even column/.append style={column sep=3pt}},
        ylabel with exponent={sig. in}{\myexpA}{\unit{\milli\volt}},
    ]
        \addplot[thick, gray] table[x index=0, y index=2] {\datafileA};
        \addlegendentry{$\signal{7}$}
        \addplot[thick, green!50!black, each nth point=10] table[x=time, y=ResidualTS] {\datafileTS};
        \addlegendentry{TS}
        \addplot[thick, red] table[x expr=\thisrowno{0}-0.112930, y index=4] {\datafileA};
        \addlegendentry{$\signal{7,\text{clean}}$}
    \nextgroupplot[
        legend style={draw=none, at={(0.95,1.35)}, anchor=north east},
        ylabel with exponent={sig. in}{\myexpB}{\unit{\milli\volt\per\cubic\second}},
    ]
        \addplot[thick, blue] table[x index=0, y index=1] {\datafileB};
        \addlegendentry{$\signalderiv{7}{3}$}
        \addplot[thick, red,dashed] table[x index=0, y index=4] {\datafileB};
        \addlegendentry{$\signalderiv{7,\text{pulse}}{3}$}
        \pgfplotstableread{\datafileD}\thresholdtable
        \pgfplotstablegetrowsof{\thresholdtable}
        \pgfmathtruncatemacro{\numthresholds}{\pgfplotsretval-1}
        \foreach \row in {0,...,\numthresholds} {
            \pgfplotstablegetelem{\row}{y_position}\of{\thresholdtable}
            \let\ypos\pgfplotsretval
            \edef\temp{\noexpand\addplot[orange, dashed, thick, domain=\pgfkeysvalueof{/pgfplots/xmin}:\pgfkeysvalueof{/pgfplots/xmax}] {\ypos};}
            \temp
        }
        \addlegendentry{Threshold}


    \end{groupplot}
\end{tikzpicture}
    \caption{Artifact removal for channel seven (male subject). Top: raw signal (gray), proposed method (red), and template subtraction (green). Bottom: derivative signal with detected pulses and threshold.}
    \label{fig:channel_7_results}
\end{figure}

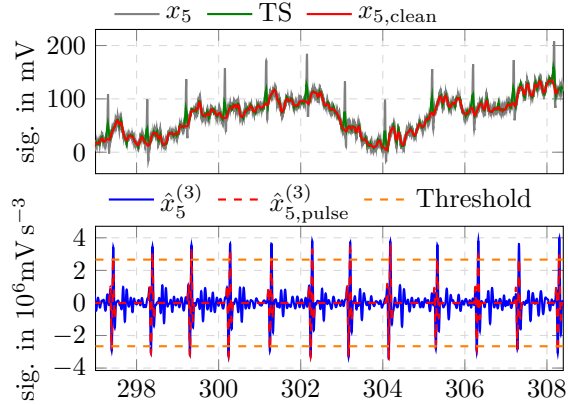
\begin{figure}
    \centering
    \begin{tikzpicture}[]
\pgfplotsset{
    ylabel with exponent/.style n args={3}{
        ylabel={#1 #3},
        scaled y ticks=true,
        ytick scale label code/.code={%
            \pgfmathparse{int(##1)}%
            \xdef#2{\pgfmathresult}%
        },
        after end axis/.code={%
            \ifx#2\undefined\xdef#2{0}\fi
            \pgfmathparse{#2}%
            \ifnum\pgfmathresult=0\relax
                \node[anchor=south,rotate=90] at (rel axis cs:-0.1,0.5) {#1 #3};
            \else
                \node[anchor=south,rotate=90] at (rel axis cs:-0.1,0.5) {#1 $10^{#2}$#3};
            \fi
        },
        ylabel style={opacity=0},
        y tick scale label style={opacity=0},
    }
}
    \def\datafileA{figs/experimental_results/csv_files_and_hash_info_male/csv_files_and_hash_info_male/channel_6_data_row0_col0.dat}
    \def\datafileB{figs/experimental_results/csv_files_and_hash_info_male/csv_files_and_hash_info_male/channel_6_data_row1_col0.dat}
    \def\datafileC{figs/experimental_results/csv_files_and_hash_info_male/csv_files_and_hash_info_male/channel_6_data_row2_col0.dat}
    \def\datafileD{figs/experimental_results/csv_files_and_hash_info_male/csv_files_and_hash_info_male/channel_6_data_row1_col0_thresholds.dat}
    \def\datafileTS{figs/experimental_results/csv_files_and_hash_info_male/csv_files_and_hash_info_male/202509010002_epi_channel_6_ecg_removed_TS.dat}

\pgfplotstableread{\datafileB}\datatable
\pgfplotstablegetrowsof{\datatable}
\pgfmathtruncatemacro{\lastrow}{\pgfplotsretval-1}
\pgfplotstablegetelem{0}{time}\of{\datatable}
\pgfmathsetmacro{\xminval}{\pgfplotsretval}
\pgfplotstablegetelem{\lastrow}{time}\of{\datatable}
\pgfmathsetmacro{\xmaxval}{\pgfplotsretval}
    \begin{groupplot}[
        group style={
            group size=1 by 2,
            vertical sep=0.7cm,
            xlabels at=edge bottom,
            xticklabels at=edge bottom,
        },
        width=7.8cm,
        height=3.5cm,
        legend pos=north east,
        legend columns=3,
        grid=major,
        grid style={dashed,gray!30},
        legend style={draw=none, at={(0.75,1.3)}, fill=none, anchor=north east, inner sep=1pt, outer sep=0pt, /tikz/every even column/.append style={column sep=3pt}},
        cycle list name=color list,
        scaled y ticks=false,
        every y tick scale label/.style={at={(yticklabel* cs:1.05)},anchor=south},
        xmin=297, xmax=\xmaxval,
    ]
    
    \nextgroupplot[legend style={draw=none, at={(0.75,1.24)}, fill=none, anchor=north east, inner sep=1pt, outer sep=0pt, /tikz/every even column/.append style={column sep=3pt}},
        ylabel with exponent={sig. in}{\myexpA}{\unit{\milli\volt}},
    ]
        \addplot[thick, gray] table[x index=0, y index=2] {\datafileA};
        \addlegendentry{$\signal{5}$}
        \addplot[thick, green!50!black, each nth point=10] table[x=time, y=ResidualTS] {\datafileTS};
        \addlegendentry{TS}
        \addplot[thick, red] table[x expr=\thisrowno{0}-0.112930, y index=4] {\datafileA};
        \addlegendentry{$\signal{5,\text{clean}}$}
    \nextgroupplot[
        legend style={draw=none, at={(0.95,1.35)}, anchor=north east},
        ylabel with exponent={sig. in}{\myexpB}{\unit{\milli\volt\per\cubic\second}},
    ]
        \addplot[thick, blue] table[x index=0, y index=1] {\datafileB};
        \addlegendentry{$\signalderiv{5}{3}$}
        \addplot[thick, red,dashed] table[x index=0, y index=4] {\datafileB};
        \addlegendentry{$\signalderiv{5,\text{pulse}}{3}$}
        \pgfplotstableread{\datafileD}\thresholdtable
        \pgfplotstablegetrowsof{\thresholdtable}
        \pgfmathtruncatemacro{\numthresholds}{\pgfplotsretval-1}
        \foreach \row in {0,...,\numthresholds} {
            \pgfplotstablegetelem{\row}{y_position}\of{\thresholdtable}
            \let\ypos\pgfplotsretval
            \edef\temp{\noexpand\addplot[orange, dashed, thick,domain=\pgfkeysvalueof{/pgfplots/xmin}:\pgfkeysvalueof{/pgfplots/xmax}] {\ypos};}
            \temp
        }
        \addlegendentry{Threshold}


    \end{groupplot}
\end{tikzpicture}
    \caption{Artifact removal for channel five (male subject). Top: raw signal (gray), proposed method (red), and template subtraction (green). Bottom: derivative signal with detected pulses and threshold.}
    \label{fig:channel_5_results}
\end{figure}

Figures~\ref{fig:channel_7_results} and \ref{fig:channel_5_results} show representative results for two channels from the male subject. The top panels display the raw signal $\signal{\chanindex}$ (gray) and the cleaned signal $\signal{\chanindex,\text{clean}}$ from the proposed method (red) and template subtraction (green) from \cite{Petersen2020} (the toolbox from \cite{Delorme2004EEGLAB} has been used for the data processing). The bottom panels show the derivative signal $\signalderiv{\chanindex}{3}$ with detected pulse signatures and the percentile-based threshold. The proposed method suppresses the cardiac artifact component while preserving signal components outside the detected pulse windows. The corresponding $\text{SNR}_{\text{out}}$ values are $\SI{5.14}{\dB}$ and $\SI{7.42}{\dB}$, respectively.

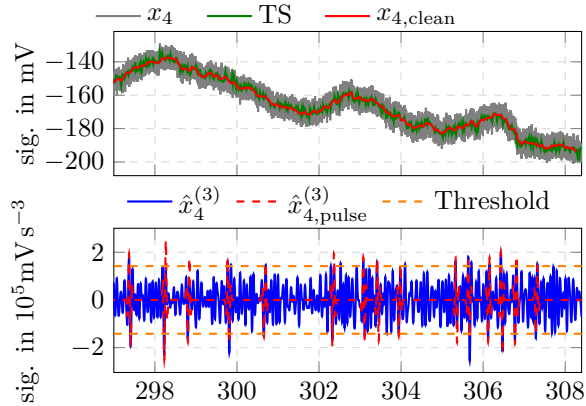
\begin{figure}
    \centering
    \begin{tikzpicture}[]
\pgfplotsset{
    ylabel with exponent/.style n args={3}{
        ylabel={#1 #3},
        scaled y ticks=true,
        ytick scale label code/.code={%
            \pgfmathparse{int(##1)}%
            \xdef#2{\pgfmathresult}%
        },
        after end axis/.code={%
            \ifx#2\undefined\xdef#2{0}\fi
            \pgfmathparse{#2}%
            \ifnum\pgfmathresult=0\relax
                \node[anchor=south,rotate=90] at (rel axis cs:-0.14,0.5) {#1 #3};
            \else
                \node[anchor=south,rotate=90] at (rel axis cs:-0.14,0.5) {#1 $10^{#2}$#3};
            \fi
        },
        ylabel style={opacity=0},
        y tick scale label style={opacity=0},
    }
}
\def\datafileA{figs/experimental_results/csv_files_and_hash_info_female/csv_files_and_hash_info_female/channel_4_data_row0_col0.dat}
    \def\datafileB{figs/experimental_results/csv_files_and_hash_info_female/csv_files_and_hash_info_female/channel_4_data_row1_col0.dat}
    \def\datafileC{figs/experimental_results/csv_files_and_hash_info_female/csv_files_and_hash_info_female/channel_4_data_row2_col0.dat}
    \def\datafileD{figs/experimental_results/csv_files_and_hash_info_female/csv_files_and_hash_info_female/channel_4_data_row1_col0_thresholds.dat}
    \def\datafileTS{figs/experimental_results/csv_files_and_hash_info_female/csv_files_and_hash_info_female/202502180003_channel_4_ecg_removed_TS.dat}
    %

\pgfplotstableread{\datafileB}\datatable
\pgfplotstablegetrowsof{\datatable}
\pgfmathtruncatemacro{\lastrow}{\pgfplotsretval-1}
\pgfplotstablegetelem{0}{time}\of{\datatable}
\pgfmathsetmacro{\xminval}{\pgfplotsretval}
\pgfplotstablegetelem{\lastrow}{time}\of{\datatable}
\pgfmathsetmacro{\xmaxval}{\pgfplotsretval}
    \begin{groupplot}[
        group style={
            group size=1 by 2,
            vertical sep=0.7cm,
            xlabels at=edge bottom,
            xticklabels at=edge bottom,
        },
        width=7.8cm,
        height=3.5cm,
        legend pos=north east,
        legend columns=3,
        grid=major,
        grid style={dashed,gray!30},
        legend style={draw=none, at={(0.75,1.3)}, fill=none, anchor=north east, inner sep=1pt, outer sep=0pt, /tikz/every even column/.append style={column sep=3pt}},
        cycle list name=color list,
        scaled y ticks=false,
        every y tick scale label/.style={at={(yticklabel* cs:1.15)},anchor=south},
        xmin=297, xmax=\xmaxval,
    ]
    
    \nextgroupplot[legend style={draw=none, at={(0.75,1.24)}, fill=none, anchor=north east, inner sep=1pt, outer sep=0pt, /tikz/every even column/.append style={column sep=10pt}},
        ylabel with exponent={sig. in}{\myexpA}{\unit{\milli\volt}},
    ]
        \addplot[thick, gray] table[x index=0, y index=2] {\datafileA};
        \addlegendentry{$\signal{4}$}
        \addplot[thick, green!50!black, each nth point=10] table[x=time, y expr=-\thisrowno{2}] {\datafileTS};
        \addlegendentry{TS}
        \addplot[thick, red] table[x expr=\thisrowno{0}-0.112930, y index=4] {\datafileA};
        \addlegendentry{$\signal{4,\text{clean}}$}
    \nextgroupplot[
        legend style={draw=none, at={(0.95,1.35)}, anchor=north east},
        ylabel with exponent={sig. in}{\myexpB}{\unit{\milli\volt\per\cubic\second}},
    ]
        \addplot[thick, blue] table[x index=0, y index=1] {\datafileB};
        \addlegendentry{$\signalderiv{4}{3}$}
        \addplot[thick, red,dashed] table[x index=0, y index=4] {\datafileB};
        \addlegendentry{$\signalderiv{4,\text{pulse}}{3}$}
        \pgfplotstableread{\datafileD}\thresholdtable
        \pgfplotstablegetrowsof{\thresholdtable}
        \pgfmathtruncatemacro{\numthresholds}{\pgfplotsretval-1}
        \foreach \row in {0,...,\numthresholds} {
            \pgfplotstablegetelem{\row}{y_position}\of{\thresholdtable}
            \let\ypos\pgfplotsretval
            \edef\temp{\noexpand\addplot[orange, dashed, thick,domain=\pgfkeysvalueof{/pgfplots/xmin}:\pgfkeysvalueof{/pgfplots/xmax}] {\ypos};}
            \temp
        }
        \addlegendentry{Threshold}


    \end{groupplot}
\end{tikzpicture}
    \caption{Artifact removal for channel four (female subject, electrodes close to uterus). Top: raw signal (gray), proposed method (red), and template subtraction (green). Bottom: derivative signal with detected pulses and threshold.}
    \label{fig:channel_5_female_results}
\end{figure}

Figure~\ref{fig:channel_5_female_results} shows results for an electrode near the uterus in the female subject. The derivative exhibits visible distortions because \change{slow-varying uterine contractions violate Assumption~\ref{ass:smooth_baseline}: the $\derivorder$-th order baseline derivative is non-zero near the uterus, producing false detections and limiting removal effectiveness.} The $\text{SNR}_{\text{out}}$ for this example is $\SI{-2.15}{\dB}$. \change{In this recording, the affected channel corresponds to the electrode positioned closest to the uterus (Figure~\ref{fig:ElectrodePlacement}), and performance is observed to degrade gradually with decreasing electrode-to-uterus distance. Whether this behavior generalizes across subjects, and whether the unaffected channels suffice for clinical monitoring, require validation on a larger cohort.}

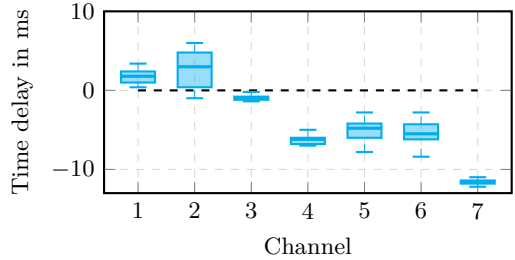
\begin{figure} 
       \centering
    \begin{tikzpicture}

    \def\datafile{figs/experimental_results/dat_files/boxplot_delays.dat}
    \begin{axis}[
        width=7cm,
        height=4cm,
        ylabel={Time delay in \unit{\milli\second}},
        xlabel={Channel},
        xtick={0,1,2,3,4,5,6},
        xticklabels from table={\datafile}{channel},
        grid=major,
        grid style={dashed,gray!30},
        tick label style={font=\small},
        label style={font=\small},
        legend style={font=\footnotesize},
        legend pos=north west,
        axis line style={thick},
        ymin = -13,
        ymax = 10,
    ]
        \addplot [black, dashed, thick, forget plot] coordinates {(0,0) (6,0)};
        
        \pgfplotstableread{\datafile}\datatable
        
        \pgfplotstablegetrowsof{\datatable}
        \pgfmathtruncatemacro{\numrows}{\pgfplotsretval-2}
        
        \foreach \row in {0,...,\numrows} {
            \pgfplotstablegetelem{\row}{index}\of{\datatable}
            \let\xpos\pgfplotsretval
            \pgfplotstablegetelem{\row}{Q1}\of{\datatable}
            \pgfmathsetmacro{\Q}{\pgfplotsretval*1000}
            \pgfplotstablegetelem{\row}{Q3}\of{\datatable}
            \pgfmathsetmacro{\Qthree}{\pgfplotsretval*1000}
            \pgfplotstablegetelem{\row}{median}\of{\datatable}
            \pgfmathsetmacro{\med}{\pgfplotsretval*1000}
            \pgfplotstablegetelem{\row}{mean}\of{\datatable}
            \pgfmathsetmacro{\mn}{\pgfplotsretval*1000}
            \pgfplotstablegetelem{\row}{whisker_min}\of{\datatable}
            \pgfmathsetmacro{\wmin}{\pgfplotsretval*1000}
            \pgfplotstablegetelem{\row}{whisker_max}\of{\datatable}
            \pgfmathsetmacro{\wmax}{\pgfplotsretval*1000}
            
            \edef\temp{\noexpand\fill[cyan, opacity=0.4] 
                (axis cs:\xpos-0.3,\Q) rectangle (axis cs:\xpos+0.3,\Qthree);}
            \temp
            
            \edef\temp{\noexpand\draw[cyan, thick] 
                (axis cs:\xpos-0.3,\Q) rectangle (axis cs:\xpos+0.3,\Qthree);}
            \temp
            
            \edef\temp{\noexpand\draw[cyan, very thick] 
                (axis cs:\xpos-0.3,\med) -- (axis cs:\xpos+0.3,\med);}
            \temp
            
            \edef\temp{\noexpand\draw[cyan, thick] 
                (axis cs:\xpos-0.15,\wmin) -- (axis cs:\xpos+0.15,\wmin);}
            \temp
            \edef\temp{\noexpand\draw[cyan, thick] 
                (axis cs:\xpos-0.15,\wmax) -- (axis cs:\xpos+0.15,\wmax);}
            \temp
            
            \edef\temp{\noexpand\draw[cyan, thin, dashed] 
                (axis cs:\xpos,\wmin) -- (axis cs:\xpos,\Q);}
            \temp
            \edef\temp{\noexpand\draw[cyan, thin, dashed] 
                (axis cs:\xpos,\Qthree) -- (axis cs:\xpos,\wmax);}
            \temp
        }
        
    \end{axis}
\end{tikzpicture}
    \caption{Inter-channel delays of detected cardiac pulses relative to channel 8.}
    \label{fig:boxplot_delays}
\end{figure}

Figure~\ref{fig:boxplot_delays} shows inter-channel delays estimated from detected cardiac pulse times relative to channel 8. Negative delays indicate earlier wavefront arrival, positive delays later arrival. Median delays range from $\SI{-11.6}{\ms}$ (channel 7) to $\SI{+3.0}{\ms}$ (channel 2), reflecting spatial propagation of cardiac electrical activity across the electrode array. The total delay spread of approximately $\SI{15}{\ms}$ corresponds to an apparent conduction velocity of $3$--$\SI{7}{\m\per\s}$ for typical electrode spacings of $5$--$\SI{10}{\cm}$, consistent with volume-conducted cardiac signals. The tight interquartile ranges ($\SI{1}{\ms}$ to $\SI{2}{\ms}$) confirm consistent detection and validate the spatial coherence assumption.

	\section{Conclusion and Outlook}\label{sec:conclusion}
An algebraic differentiator-based approach for removing cardiac artifacts from EHG signals has been presented. Differentiation enhances impulsive cardiac signatures, enabling detection via percentile thresholding and pulse reconstruction. Experimental results demonstrate effective removal without an ECG reference, with consistent inter-channel delays validating the spatial coherence assumption.

 Future work should extend the signal model to accommodate non-zero baseline derivatives and \change{extend validation to a larger cohort including subjects with active uterine contractions, carry out quantitative benchmarking using a metric applicable to all compared methods, conduct systematic parameter sensitivity analysis across subjects and recording conditions, and investigate clinical performance metrics such as detection sensitivity and specificity as objectives for differentiator parameter selection.}
	
	\bibliographystyle{IEEEtran}
	\bibliography{bib}
\end{document}